# Sensing supercurrents using geometric effects

Adam N. McCaughan[†], Nathnael Abebe[†], Qing-Yuan Zhao[†], Karl K. Berggren[†]

[†] Research Laboratory of Electronics, Massachusetts Institute of Technology,

77 Massachusetts Avenue, Cambridge, Massachusetts 02139, USA.

Traditional superconducting electronic devices have exploited nonlinear two-terminal devices based on tunnel junctions or nanowires for photodetection, magnetic-field sensing, and large-scale classical and quantum computing [1–4]. Because of the inconvenience of sensing and amplifying signals with a two-terminal device, researchers have tried to develop a number of three-terminal superconducting devices [5–8] based on a variety of microscopic physical mechanisms, including integration with thermal [9,10], semiconducting [11], and magnetic [12–14] systems.  However, existing three-terminal device proposals have had two main problems that have prevented their wide adoption: (1) they required the combination of superconducting materials with tunnel barriers or complex materials e.g. semiconducting or magnetic materials, which led to significant engineering challenges in materials growth, device design, and device fabrication; (2) during operation, they typically induced a voltage in series with the supercurrent carrying-wire, thus interfering with and disturbing the signal to be sensed.  These problems have prevented the broad applicability of these proposed devices. Here we describe a superconducting three-terminal device that uses a simple geometric effect known as current crowding[15] to sense the flow of current and actuate a readout signal. The device consists of a "Y"-shaped current combiner, with two currents (sense and bias) entering through the top arms of the "Y", intersecting, and then exiting through the bottom leg of the "Y'". This geometry--mixing two inputs at a sharp intersection point--takes its inspiration from Y-shaped combiners in fluid flow systems, where variations in the input pressures can produce at turbulence and mixing at the intersection [16–18]. When current is added to or removed from one of the arms (the sense arm), the superconducting critical current in the other arm (the bias arm) is modulated. The current in the sense arm can thus be determined by measuring the critical current of the bias arm. The dependence of the bias critical current on the sense current is possible because current crowding causes the sense current to interact locally with the bias arm. Measurement of the critical current in the bias arm does not break the superconducting state of the sense arm or of the bottom leg, and thus the signal to be sensed is fully restored after the measurement process. This device thus has potential for broad applicability across superconducting technologies and materials.

In the yTron, the sense supercurrent can be measured nondestructively because the current crowding effect directly modifies the Gibbs free energy barrier of the bias arm [19,20]. Previously it has been possible to measure supercurrents nondestructively, but typically these techniques involved coupling to an induced magnetic field like in the SQUID amplifier [21,22] or measurement of a macroscopic material parameter which is modified by the supercurrent [23,24,1,4]. Although topologically distinct, this device bears closest resemblance to the superconducting low-inductance undulatory galvanometer (SLUG) microwave amplifier [25,26]. The SLUG is a relative of the SQUID amplifier in which the phase of the readout SQUID was shifted by galvanically injected currents rather than a coupled magnetic field. The SLUG was capable of producing quantum-limited amplification, but like the SQUID required fabrication of two Josephson junctions.

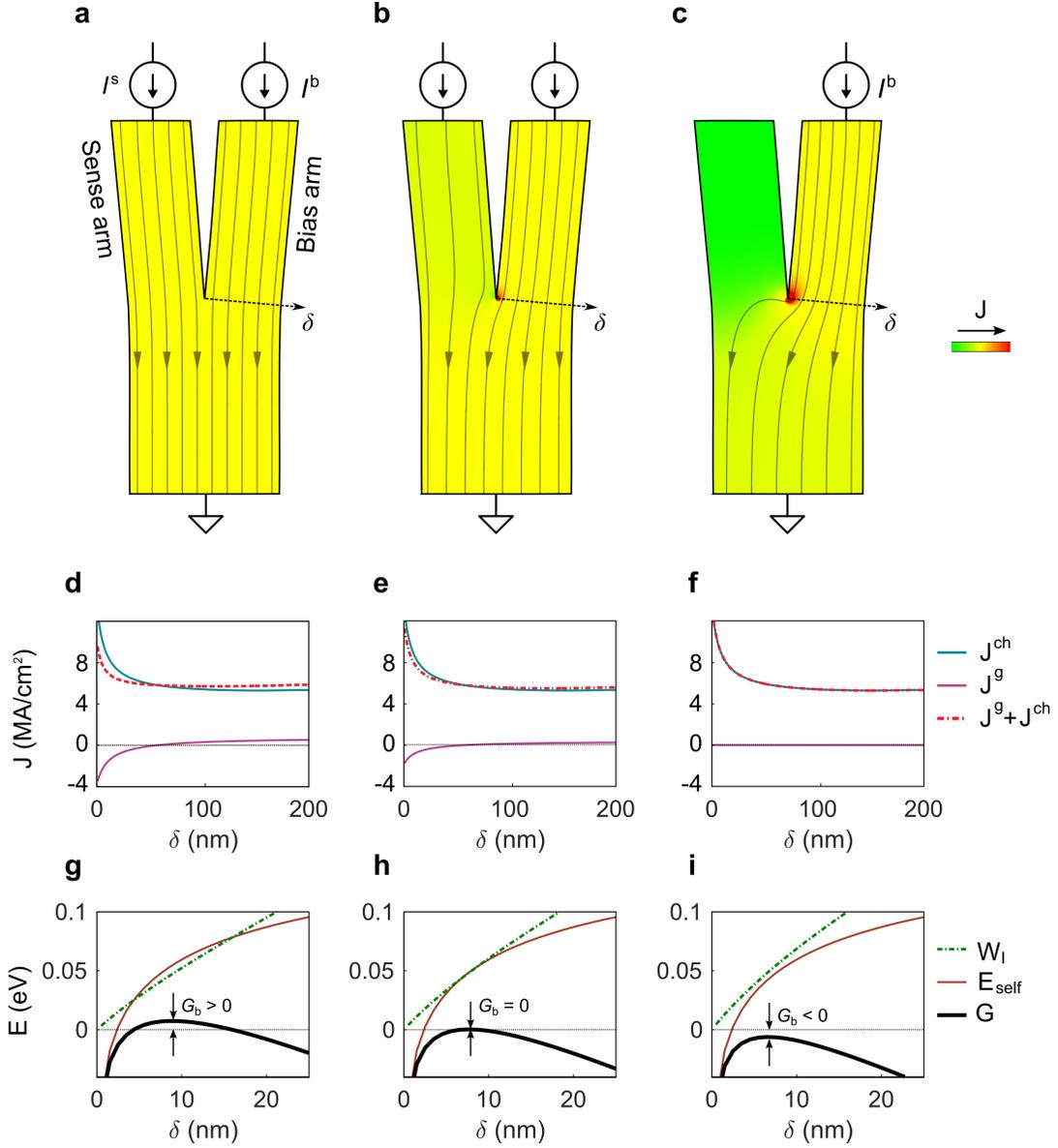

**Figure 1.** Illustration of the yTron geometric current crowding and its effect on the Gibbs free energy barrier to vortex entry. (a-c) The sense arm is biased with the same current density as the bias nanowire, and there is minimal current crowding at the intersection. The current contributions yield a current density $J$ that is mostly uniform in the bias arm, and the resulting vortex energy barrier $G_b$ is well above zero. (d-f) The sense current $I^s$ is reduced and current begins to crowd on the bias arm side in the vicinity of the intersection, reducing $G_b$. (g-i) $I^s$ is set to zero, and as a result the streamlines in the bias arm curve sharply around the intersection,

indicating significant current crowding. Under these bias conditions, $G_b$ is reduced below zero and flux (vortices) flows across the bias arm.

The functionality of the yTron arises from the dependence of the current distribution in the bias arm on the sense current--unlike most superconducting devices, it requires no tunnel junctions or loops. Due to current crowding, supercurrent flowing into the sense arm "detours" briefly into the bias arm region as it flows towards the source terminal. This detour produces zero net current flux into the bias arm, but it does modify the local distribution of current density along the width of the bias arm. Near the intersection especially, the current density in the bias arm depends strongly on the magnitude of the sense current. Ultimately, the modification of the current density by the sense current produces changes in the Gibbs free energy barrier to vortex entry along the width of bias arm, either raising or lowering the barrier to vortex entry. As a result, although the yTron geometry is fixed, the effective critical current of the bias arm can be increased or decreased by adding or removing current from the sense arm much like the SQUID or SLUG microwave amplifier.

We can show the impact of sense-current-induced current crowding on the bias critical current using the three example conditions shown in Fig. 1. In each of these examples, we computed the current flow through the device, then followed the methods described in Ref. 12 to calculate the vortex self-energy $E_\text{self}$ along with the work done by the current sources $W_I$ at each point in space. These contributions were then summed to create a map of $G$, the Gibbs free energy barrier to vortex entry and ultimately determine the height of that barrier in the bias arm, $G_b$. A slice of this map along the bias arm transverse axis $\delta$ is shown in Fig. 1 to facilitate understanding (examples of the complete map are shown in Fig. 2(d)) for each of the example bias conditions. In Fig. 1(a,d,g), the yTron is biased such that the current densities through the sense and the bias arms are equal. The current streamlines from both arms combine uniformly at the intersection and flow to the source, which corresponds to a minimally-crowded state. The current density $J$ along the bias arm ($\delta$ axis) is maximally uniform and results in a $G_b$ is greater than zero, which prevents vortices from crossing and creating a voltage state in the bias arm. In Fig. 1(b,e,h), the current flowing into the sense arm has been reduced, and as a result the current streamlines from the bias arm bend slightly around the intersection point on their way to the source terminal. Under these conditions, the the changes in $J$ near the intersection lead to an increase in the the work done on the vortex $W_I$, such that the energy barrier $G_b$ is equal to zero. In Fig. 1(c,f,i), the sense current has been shut off, and the remaining streamlines bend sharply around the intersection point indicating significant current crowding. Correspondingly, $J$ and $W_I$ increase in magnitude near the intersection. This increase is responsible for driving $G_b$ below zero, and for these bias conditions we would see the generation of a voltage state in the bias arm--either in the form of vortex crossings [27] or if underdamped, as a macroscopic hotspot [28]. This voltage appears exclusively between the bias and source terminals: $G_b$ remains greater than zero in the sense arm and so its superconducting state is unperturbed.

To demonstrate and validate the operation of the yTron, we fabricated and characterized several devices of varying dimensions. Fig. 2(a-d) shows the characterization of a yTron with 200-nm-wide arms. As expected, the IV curve of the bias arm was electrically identical to that of a hysteretic superconducting nanowire [29], albeit one that has an critical current dependent on the sense current. As shown in Fig. 2(c), the dependence of the bias arm critical current $I_c$ on the sense input $I^s$ was approximately linear over a large range of $I^s$ values and had a slope of 0.52 with an intercept of 49.3 µA. The width of the $I_c$ transition was 0.53 µA, as measured at full bandwidth a 6 GHz oscilloscope. We produced a theoretical fit of $I_c(I^s)$ by calculating $G$ everywhere in the geometry given value of $I^s$, extracting the minimum barrier location $G_b$ from the map, then calculating how much bias current was required to reduce $G_b$ to zero. For parameters, we calculated that the nanowires had a Pearl length $\Lambda = 2\lambda^2/d$ of 85 µm, where $d$ was the thickness of the NbN film which was measured to be 4.8 nm, and a penetration depth $\lambda$ was set to be 450 nm [30–32]. The best fit to the data was found with when the radius-of-curvature of the intersection, $\rho_c$, was set to 5 nm, which matched the approximate fabrication resolution of the HSQ-based electron-beam process used to pattern the device [33]. The theoretical fit diverged from the empirical results below $I^s$ = -7 µA; this behavior is consistent with the vortex formation changing from vortex nucleation point at the intersection ($\delta$ = 0 nm) to antivortex nucleation the far edge ($\delta$ = 200 nm). We note that although we fit $\rho_c$ to the data, as shown in Fig. 2(c) there is only a weak dependence of $I_c(I^s)$ on $\rho_c$, suggesting that device operation is tolerant to variations and errors in the fabrication process. We were able demonstrate this tolerance by characterizing 22 different devices with arm widths $w_s$ and $w_b$ between 100 nm and 800 nm as shown in Fig. 2(e). The dependence of $I_c$ for the bias arm depends primarily on the current density at the intersection (determined by $I_s$ and $w_s/w_b$), and weakly on $\rho_c$.

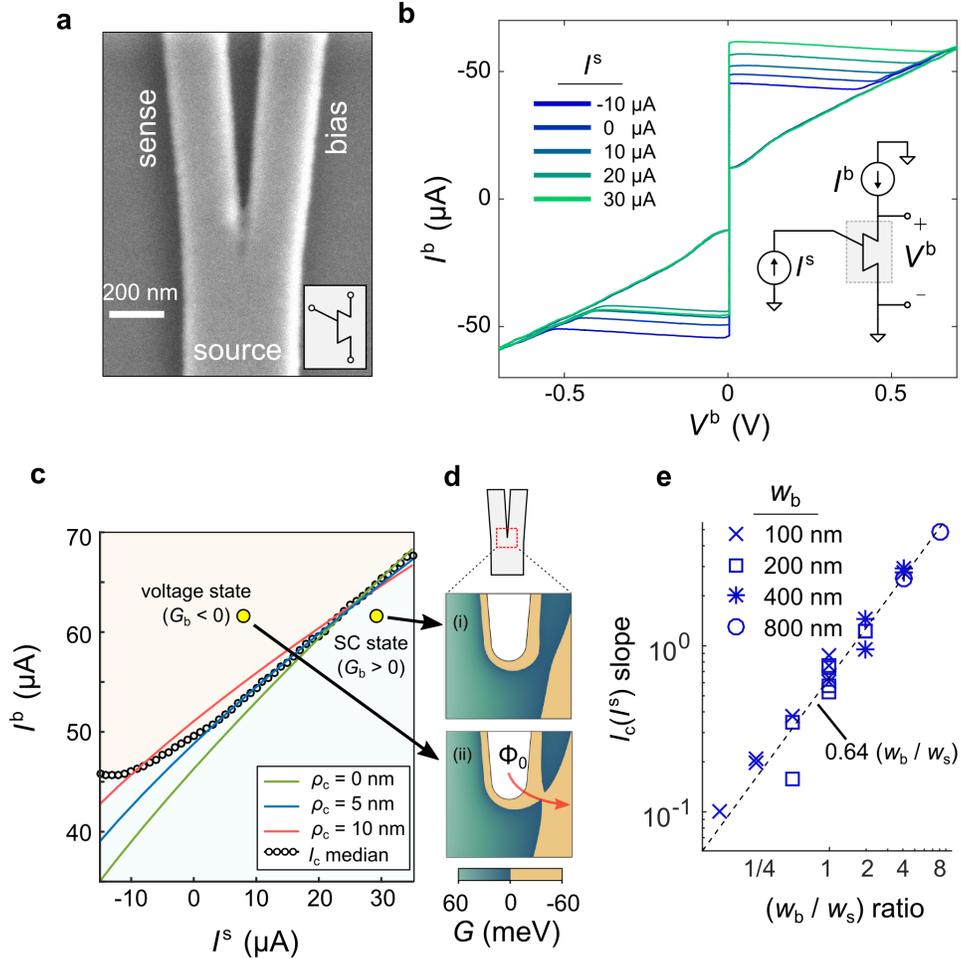

**Figure 2.** Basic operation of the yTron. (a) Scanning electron micrograph of a yTron with 200-nm-wide arms (b) I-V characteristics of the bias arm for five values of $I^s$. (inset) Circuit schematic for testing the yTron at different $I^s$ bias points. (c) Plot of the yTron regimes of operation for input conditions $I^s$ and $I^b$. The border between the conditions $G_b>0$ and $G_b<0$ defines $I_c(I^s)$, and is denoted by the black dots (experimental data) and the red, blue and green lines (theory plots). (d) Computation of the Gibbs free energy map for vortex entry illustrating the transition between the superconducting and voltage states by changing $I^s$. (i) $G_b>0$ and there is no path for vortices to enter and cross the bias arm (ii) When $I^s$ is reduced, a gap opens in the barrier landscape so that there is a continuous path across the bias arm where $G<0$; vortices will flow and cause a voltage to appear in the bias arm. (e) Log-log plot of the experimentally-measured slopes of $I_c(I^s)$ versus geometric arm width ratio for 22 different devices.

There are several aspects of the yTron design which impact its operation. The first characteristics which must be considered are those of the superconducting material from which the device is fabricated. The superconducting film thickness must be less than the material's penetration depth $\lambda$ in order for the current crowding effect to work as described. In a thicker superconductor with a non-uniform kinetic inductance, current may not be distributed evenly across the cross-section of each arm of the yTron, altering the effect of current crowding. By making the device from a film thinner than $\lambda$ and with arm widths less than $\Lambda$, the device has an (approximately) uniform sheet kinetic inductance that produces the current distributions shown in Fig. 1. Similarly, the arms of the yTron should be wider than the coherence length (in thin-film NbN $\xi \approx 4$ nm) [34]. Arms with widths on the order of the coherence length are effectively one-dimensional and may not exhibit the required current-crowding effects. Ultimately, the dependence of the bias critical current $I_c$ on the sense current $I^s$ is based on three geometric elements: (1) the "sharpness" of the intersection point, (2) the widths of the bias and sense nanowires, and (3) the angle at which the sense and bias arms intersect. In an ideal Ohmic conductor, an infinitely sharp point at the intersection tip would result in a diverging current density for any streamline bending around the intersection. However, in the sharpness of the intersection point is ameliorated by two factors: rounding caused by the practical fabrication limits of e-beam lithography, and the superconducting radius-of-curvature effect (as described in Ref. 12) which produces a rounding of the intersection point on the order of the material superconducting coherence length, even for a perfectly sharp intersection.

Isolation between the sense and bias is a key feature of the yTron. Since the yTron is fabricated from a continuous superconducting film, "isolation" in this context does not mean capacitively decoupled like a FET, but instead the isolation of the sense current from changes in the bias arm. Even when the superconductivity in the bias arm breaks down--e.g. a voltage state forms--the superconducting state of the sense nanowire is not disrupted. Due to the indirect nature of the current-crowding-based modulation of $I_c$, a voltage state in the bias (source to bias) does not produce any voltage on the sense (sense to bias).

As an example, let us assume we have a yTron which is biased just below $I_c$, as in Fig. 2(d)(i). When $I^s$ is reduced, the barrier $G_b$ will be reduced and vortices will begin to flow across the bias arm as shown in Fig. 2(d)(ii). The flow of vortices will produce a voltage between the source and bias terminals. Despite the fact that flux is passing across the bias nanowire, the superconducting state in the sense nanowire never broke down, and no flux was able to cross. One concern we had when testing the yTron was that excited quasiparticles generated by the hotspot or vortex crossings could diffuse and cause a breakdown of superconductivity in the sense arm. In thin-film NbN, this diffusion length is ~100 nm [35], on a similar scale to the nanowire widths. However, for the device shown in Fig. 1, we found that as long as the total power dissipation caused by the voltage in the bias arm was below 350 nW the sense arm was not disrupted.

Due to its ability to measure inline supercurrent, a natural application for the yTron is the readout of quantized currents in a superconducting loop. By placing the sense of the yTron inline with a superconducting loop, we were able to use the yTron to nondestructively read out the number of discrete fluxons ($n$) trapped in a the loop, as shown in Fig. 3. We successfully resolved 13 adjacent fluxon states ($n$, $n+1$, etc) of the loop, and were able to read out those states several thousand times consecutively without changing the value of $n$. This application was possible because the sense (loop) supercurrent can be inferred from $I_c$, and can be measured repeatedly without changing $n$ by allowing flux into or out of the loop.

Using the circuit shown in Fig. 3(a), we observed the quantization of current in the loop with an experiment which consisted of two alternating steps. These steps are depicted in Fig. 3(b). First we performed a readout experiment in which we measured $I_c$ with 100 consecutive trials. Each trial consisted of steadily increasing $I^b$ up from zero until $I_c$ was reached, indicated by a ~5 mV voltage appearing in the bias arm. After these trials, we then performed a "write" in which we applied an external voltage pulse to the loop to deliberately break the superconducting loop temporarily and allow the number of stored fluxons *n* to change randomly. Repeating this two-step process several times, we then plotted the median $I_c$ value and standard deviation of the distributions produced by each readout experiment, shown in Fig. 3(c).

The results shown in Fig. 3 indicate that the supercurrent in the yTron sense arm was isolated from the breakdown of superconductivity in the bias arm, and that the input supercurrent $I^s$ could be measured without changing its value. We found that only the write process was able to change the median $I_c$, and we additionally found that the median $I_c$ shifted in discrete steps of 2.61 µA. The steplike jumps of the $I_c$ indicated that individual fluxons in the loop were being added ($n \rightarrow n+1$) or subtracted ($n \rightarrow n-1$) by the write process, the hallmark of fluxoid quantization in superconducting loops [36]. The addition or removal of a fluxon to the loop changed $I^s$ by a quantized amount $\Phi_0/L$ which in turn shifted the approximately-linear $I_c(I^s)$ by a steplike amount.

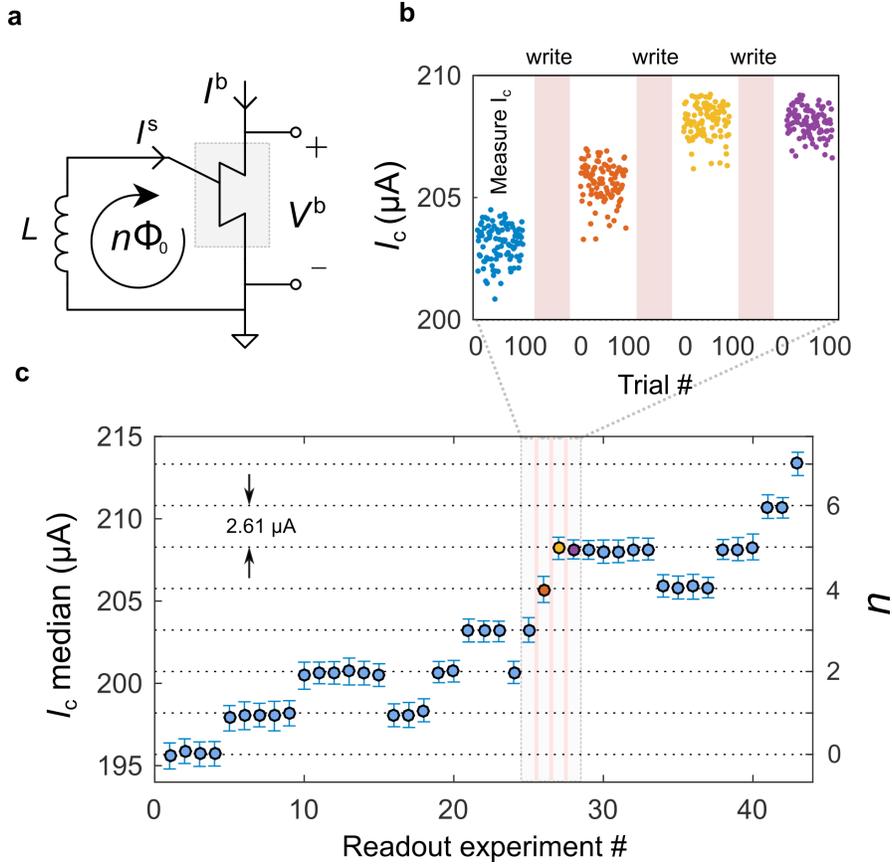

**Figure 3.** Resolving individual fluxons in a closed superconducting loop using the yTron as an inline current sensor for a superconducting loop. (a) Superconducting circuit used to trap $n$ fluxons and pass the quantized current $n\Phi_0/L$ into the yTron sense arm. (b) Readout measurements of the bias arm critical current $I_c$. After each set of 100 trial measurements of $I_c$, a write operation was performed which allowed $n$ to change, and then another set of 100 measurements was taken. (c) Plot showing the median $I_c$ (dots) and standard deviation (bars) of each set of 100 trials and their correspondance to the fluxon state $n$; between each experiment was a write operation which added or removed fluxons randomly. The steplike change in the $I_c$ median demonstrates that the yTron resolved adjacent fluxon states of the persistent loop current. The uniformity of the $I_c$ median values indicates that during the readout process $n$ remains unchanged, even though a voltage $V_b$ forms in the bias.

In conclusion we have developed, characterized, and applied the yTron, a new three-terminal superconducting device which is able to sense superconducting currents inline without perturbing them, has a nanoscale active area, and is insensitive to and fabrication variations. The yTron has immediate applications as an inline current sensor for devices such as transition edge sensors and superconducting nanowire single photon detectors (SNSPD), and as a compact superconducting memory. Of further interest is its relation to the SQUID and microwave SLUG amplifiers. The yTron shares enough operational features with these devices to suggest it may be used similarly, as a sensitive superconducting amplifier by damping the bias arm with a shunt resistor. At the same time, the yTron geometry is naturally sub-micron, and its topology means many of the typical fabrication barriers facing superconducting electronics (such as tunnel-barrier fabrication) are avoided entirely. Since the yTron functionality comes from current crowding which occurs in every superconductor, it should be possible to fabricate it from any superconducting material, even 2D superconductors like $NbSe_2$ or proximitized graphene. Additionally, the form of the yTron lends itself well to a CMOS-type fabrication scheme: its geometry can also be reproduced vertically by using an oxide layer between two superconducting thin films as the barrier between the sense and bias arms. Another potential use of the yTron would be to use it as a three-terminal controllable weak link in the style of ultrafast, low-power single-flux quantum logic like RSFQ [37,38]--by forming loops between the sense and bias and source and bias, flux flow into the source-bias could be controlled by the motion of flux into and out of the sense-bias loop.

**Acknowledgements:** The authors would like to James Daley and Mark Mondol for nanofabrication technical support. This work was supported by the AFOSR  Adam McCaughan was supported by a fellowship from the NSF iQuISE program, award number 0801525.